# Tripling energy storage density through order-disorder transition induced polar nanoregions in PbZrO$_3$ thin films by ion implantation

Yongjian Luo[1], Changan Wang[2,3], Chao Chen[1], Yuan Gao[4], Fei Sun[1], Caiwen Li[1], Xiaozhe Yin[1], Chunlai Luo[1], Ulrich Kentsch[3], Xiangbin Cai[5], Mei Bai[6], Zhen Fan[1], Minghui Qin[1], Min Zeng[1], Jiyan Dai[7], Guofu Zhou[8], Xubing Lu[1], Xiaojie Lou[6], Shengqiang Zhou[2], Xingsen Gao[1], Deyang Chen[1,a)], Jun-Ming Liu[1,9]

[1]Institute for Advanced Materials and Guangdong Provincial Key Laboratory of Optical Information Materials and Technology, South China Academy of Advanced Optoelectronics, South China Normal University, Guangzhou 510006, China

[2]Helmholtz-Zentrum Dresden-Rossendorf, Institute of Ion Beam Physics and Materials Research, Dresden 01328, Germany

[3]Guangdong Institute of Semiconductor Industrial Technology, Guangdong Academy of Sciences, Guangzhou 510650, China

[4]State Key Laboratory of Nuclear Physics and Technology, School of Physics, Peking University, Beijing 100871, China

[5]Division of Physics and Applied Physics, School of Physical and Mathematical Sciences, Nanyang Technological University, Singapore 637371, Singapore

[6]Frontier Institute of Science and Technology, State Key Laboratory for Mechanical Behavior of Materials, Xi'an Jiaotong University, Xi'an, 710049 China

[7]Department of Applied Physics, The Hong Kong Polytechnic University, Hung Hom, Kowloon, Hong Kong, China

[8]National Center for International Research on Green Optoelectronics, South China Normal University, Guangzhou 510006, China

[9]Laboratory of Solid State Microstructures and Innovation Center of Advanced Microstructures, Nanjing University, Nanjing 210093, China

a)Authors to whom correspondence should be addressed: deyangchen@m.scnu.edu.cn

# ABSTRACT

Dielectric capacitors are widely used in pulsed power electronic devices due to their ultrahigh power densities and extremely fast charge/discharge speed. To achieve enhanced energy storage density, both maximum polarization ($P_{max}$) and breakdown strength ($E_b$) need to be improved simultaneously. However, these two key parameters are inversely correlated. In this study, order-disorder transition induced polar nanoregions (PNRs) have been achieved in $PbZrO_3$ thin films by making use of the low-energy ion implantation, enabling us overcome the trade-off between high polarizability and breakdown strength, which leads to the tripling of the energy storage density from 20.5 $J/cm^3$ to 62.3 $J/cm^3$ as well as the great enhancement of breakdown strength. This approach could be extended to other dielectric oxides to improve the energy storage performance, providing a new pathway for tailoring the oxide functionalities.

## I. INTRODUCTION

Dielectric capacitors, exhibiting ultrafast charging and discharging rates, high voltage endurance and good reliability, are promising for applications in integrated circuits and modern energy storage devices [1-4]. In electrostatic capacitors, electrical energy can be stored and released by the polarization and depolarization upon application and removal of electrical field. The recoverable energy density, $U_{re}$, can be extracted from the polarization–electric field (P-E) hysteresis loops by equation $U_{re} = \int_{P_{rem}}^{P_{max}} EdP$, where E is the applied electric field, $P_{max}$ is the maximum polarization and $P_{rem}$ is the remanent polarization [5]. Consequently, the achievement of high $P_{max}$, low $P_{rem}$ and large breakdown strength ($E_b$) are critical to obtain high performance electrostatic capacitors. However, dielectric materials with large polarization usually present high dielectric constant while the breakdown strength usually decreases as the dielectric constant (k) increases as demonstrated by $E_b = k^{-0.65}$ [6,7]. Therefore, overcoming the trade-off between $P_{max}$ and $E_b$ is necessary to achieve high performance dielectric capacitors.

Recently, the design of ferroelectric domain at nanoscale has attracted a lot of attention in energy storage dielectrics [8,9]. Compared to micrometer-scale domains in typical ferroelectrics, these microscopic polar nanoregions (PNRs) exhibit lower energy barriers for polarization switching, accordingly reducing the energy dissipation caused by domain wall motion and giving rise to the optimized energy density and efficiency [10]. The preparation of oxide solid solutions by complex composition engineering to construct PNRs have achieved significant success [11-13]. Combining ferroelectrics with paraelectric phase, the long-range ferroelectric order is decomposed into morphotropic or polymorphic nanodomains, such as $BaTiO_3$-$BaZrO_3$ [14], $PbMg_{1/3}Nb_{2/3}O_3$-$PbTiO_3$ [15] and $BiFeO_3$-$SrTiO_3$ [16]. In addition, chemical doping [17] and strain engineering [18] are also effective methods to induce PNRs in typical ferroelectrics. However, composition-driven polymorphic domain design would sacrifice the polarization, chemical doping may form a complex multivalent system, and strain engineering is limited by films thickness and the lack of commercial single-crystal substrates.

Ion beam technique has been used to successfully modify the structures and composition of multifunctional oxides [19,20]. Broad properties such as ferroelectricity[21], dielectricity[22], magnetism[23] and metal-insulator transition[24] can be controllably tuned by the species, energy and fluence of implanted ions. Small-volume implanted ions, such as helium, can enter the interstitials of the lattices under low-energy implantation, and cause lattice distortion or even amorphization, which may be beneficial for breaking the long-range ferroelectric order and constructing PNRs. Recent study shows that the relaxor ferroelectric (RFE) $0.68Pb(Mg_{1/3}Nb_{2/3})O_3$-$0.32PbTiO_3$ (PMN-PT) exhibits great improvement of both high-field polarizability and breakdown strength after the high-energy helium ion implantation[25], demonstrating the great application potential of ion implantation for high-performance energy storage dielectric. However, the effects of low-energy ion implantation on the energy storage performance of antiferroelectric materials remains elusive.

In this work, we focus on the model antiferroelectric $PbZrO_3$ (PZO), which shows long-range ordered antiparallel arrangement of dipoles[26]. Through the low-energy helium ion implantation induced order-disorder transition in PZO thin films, PNRs can be induced which enable us to overcome the trade-off between high polarizability and breakdown strength. The ground state AFE orthorhombic symmetry of PZO thin film can be driven to a new tetragonal phase by He ion implantation. A combination of polarization–electric field hysteresis loop (P-E loop), piezoresponse force microscopy (PFM) and scanning transmission electron microscopy (STEM) measurements indicate that it is an ion implantation driven AFE to RFE-like (order-disorder) phase transition. The PNRs induced by ion implantation leads to simultaneously tripling the energy storage density from 20.5 J/cm$^3$ to 62.3 J/cm$^3$ and greatly enhancing the breakdown strength in PZO thin films.

## II. RESULTS AND DISCUSSION

50-nm-thick PZO thin films were grown on (001)-orientated $SrTiO_3$ (STO) single crystal substrates with 50-nm-thick $La_{0.7}Sr_{0.3}MnO_3$ (LSMO) as bottom electrodes via pulsed laser deposition. Figure 1 shows the X-ray diffraction (XRD) $\theta$–$2\theta$ scans and reciprocal space mappings (RSMs) of the as-grown PZO thin film and Helium ion implanted PZO films with various doses. The as-grown PZO thin film presents typical orthorhombic perovskite crystal structure (*Pbam,* a=5.87 Å, b=11.74 Å, c=8.20 Å, where a, b, c are the lattice constant[27]) with two diffraction peaks of (240)$_O$ and (004)$_O$ ("O" denotes orthorhombic indices), which results from the large lattice mismatch between PZO ($a_{pc}$=4.16 Å, where "pc" refer to the pseudocubic unit cell) and LSMO ($a_{pc}$=3.874 Å) [28,29]. With the increase of He ion implantation doses, both (240)$_O$ and (004)$_O$ peaks gradually shift and finally emerge to lower 2θ values (Fig.1(a)), suggesting larger d-spacings induced by ion implantation. RSMs around the (440)$_O$ reflection, corresponding to the coexistence of 90° structural domains in PZO[30], are shown in Fig. 1(b). The as-grown PZO as well as the implanted samples with the dose of 2.5×10$^{14}$ and 5×10$^{14}$ ion/cm$^2$ display (440)$_O$ and (126)$_O$ diffraction patterns, in consistent with the (240)$_O$ and (004)$_O$ peaks in *θ-2θ* line scan displayed in Fig. 1(a). Higher He implanted doses of 2.5×10$^{15}$ and 5×10$^{15}$ ion/cm$^2$ lead to single-peak dominant diffraction spot, which reveals the same in-plane lattice constants of 4.13 Å while expanded out-of-plane lattice constants from 4.16 Å to 4.18 Å with the increase of implanted doses.

Further RSMs around the (103) reflection of PZO films on STO substrates were performed to confirm the phase transition driven by He ion implantation (Fig. 2 a, b and Fig. S1). In the antiferroelectric PZO, the frozen Σ mode that gives rise to the antiparallel displacement of the lead atoms exhibits 2π/a(1/4,0,1/4) quarter-order diffraction peaks in a pseudocubic unit cell, where a is the pseudocubic unit cell lattice constant[31,32]. In Fig. 2 (a) and (b), the main PZO diffraction patterns are indexed as (440)$_O$, while the quarter-order Bragg diffraction patterns from the antiparallel displaced lead atoms are indexed as (450)$_O$ and (430)$_O$. The RSM data of the as-grown PZO film show both the main and quarter-order diffraction patterns, revealing the existence of the typical antiferroelectric order. In contrast, RSM results of the implanted PZO (with the dose of 5×10$^{15}$ ions/cm$^2$) present considerably reduced intensity of the quarter-order Bragg peaks. Line profiles along the L-direction of (430)$_O$ diffraction pattern in Fig. 2(c) show two orders of magnitude lower diffraction intensity of the implanted PZO comparing with the as-grown sample, further confirming the sharp decrease of the antiferroelectric phase.

To further study the structure of implanted PZO thin films, the cross-sectional

STEM image of implanted PZO with the dose of $2.5\times10^{15}$ ion/cm$^2$ is displayed in Fig. 2 (d). The sharp interface between the PZO and LSMO layer confirm the high quality of the films. The fast Fourier transform (FFT) shown in Fig. 2(e) obtained from the top dashed rectangle region reveals the tetragonal symmetry of the implanted PZO. However, the FFT of the bottom dashed rectangle region near the PZO/LSMO show the ¼ <110> superlattice reflections (the red squares in Fig. 2(f)) due to the existence of the antiferroelectric phase as displayed in the blue dashed circle in Fig. 2. This feature and the 90° domain wall in Fig. S2 indicate the existence of a small amount of orthorhombic phase. These results are consistent with the RSM data in Fig.2(a-b). This can be explained as the He ions depth distribution follow the Gaussian distribution, thus the concentration of He ions in the region close to the interface was not enough to induce phase transition. A combination of XRD, RSM and STEM data in Fig. 1 and Fig.2 demonstrate the orthorhombic (antiferroelectric) phase to tetragonal phase transition induced by He ion implantation.

Next, polarization–electric field (P-E) hysteresis loop measurements were carried out to study the ferroelectric behaviors of the as-grown and implanted PZO thin films. As shown in Fig. 3(a), the P-E hysteresis loop of the as-grown PZO exhibits the typical antiferroelectric double loops. Interestingly, He ion implantation leads to the gradual change of the antiferroelectric behavior with the increase of implanted doses in PZO from $2.5\times10^{14}$ ion/cm$^2$ (Fig. 3(b)) to $5\times10^{14}$ ion/cm$^2$ (Fig. 3(c)). Further implanted doses give rise to the disappearance of antiferroelectricity (Fig. 3(d) and (e)). The single P-E hysteresis loops become slimmer and both remanent polarization and coercive field are reduced. The evolution of double P-E hysteresis loop to single slim hysteresis points to the AFE to RFE-like (order-disorder) transition induced by ion implantation in PZO thin films. The reversible polarization switching, phase loop and butterfly loop further confirm the ferroelectricity of the implanted PbZrO$_3$ (Fig. S8).

Among different kinds of dielectric materials, antiferroelectrics and relaxor ferroelectrics have strong application potentials for high-energy storage capacitors. However, antiferroelectrics usually show low saturation polarization while relaxor ferroelectrics have relatively low breakdown field, inhibiting the enhancement of energy storage densities[33]. Here, the He ion implantation induced AFE to RFE-like transition enables the enhancement of saturation polarization and the improvement of breakdown field simultaneously in PZO thin films. As shown in Fig. 4(a), the implanted PZO (with the He implantation dose of $2.5\times10^{15}$ ion/cm$^2$) possesses much larger saturation polarization and higher breakdown strength than the as-grown sample. The corresponding energy storage densities are shown in Fig. 4(b) and Fig. S3. Compared to the energy storage density of 20.5 J/cm$^3$ in the as-grown PZO, the energy storage density in the implanted PZO with the dose of $2.5\times10^{15}$ ions/cm$^2$ can be enhanced to 62.3 J/cm$^3$. The decrease of the energy density in higher doses of the implanted PZO may be aroused by the emergence of the amorphization phase indicated from the decrease of the diffraction intensity of implanted PZO with the increase of injection doses from $2.5\times10^{15}$ to $5\times10^{15}$ ion/cm$^2$, as shown in Fig. 1(b). Furthermore, we obtained statistical values of breakdown field strength ($E_{BDS}$) by testing 8 capacitors to failure, and fitted the distribution to a standard Weibull distribution (Fig. 4(c) and Fig. S3(d)). It is found that the characteristic $E_{BDS}$ (and Weibull modulus β, which represents the dispersion in the data) are 3.087 MV/cm (β=5.4) and 4.493 MV/cm (β=5.7) for the as-grown and the implanted PZO, respectively. The enhanced $E_{BDS}$ of the implanted PZO allows both the improvement of working reliability and energy density of dielectric capacitors. Previously reported methods, such as chemical doping[34-36], multilayer design[37], were used to improve the energy storage density of PZO. The energy density can be enhanced to ~30 J/cm$^3$. Very recently, flexible PZO thin films with high energy density of 46~ 52 J/cm$^3$ were achieved on muscovite substrate[38]. Compared with these studies, the energy storage density of 62.4 J/cm$^3$ reported in our work is competitive and the ion implantation technique is compatible with semiconductor industry, which has broad application prospects. The frequency and temperature dependence of the P-E loops (Figure S7) show good stability of our samples.

To further understand the mechanism of the enhancement of energy storge performance in the implanted PZO samples, STEM and PFM measurements (Fig. 5 and Fig. S4-S5) were carried out on the He implanted PZO sample with the dose of $2.5\times10^{15}$ ion/cm$^2$. Determined by Zr$^{4+}$ relative to the lattice center of its four nearest neighboring Pb$^{2+}$, the polarization vector mapping in Fig. 5a, based on the original STEM image in Fig. S5, shows different polarization direction nanodomains with the size of 5-10 nm, indicating the formation of PNRs in the implanted PZO thin film. The out of plane (OOP) PFM amplitude and phase images further confirm the achievement of high-density nanodomains. It is worth to mention that we use the AC voltages of 0.5 V to acquire the data to avoid the evolution of the nanodomains. These polar nanodomains is consistent with the characteristics of relaxor-like ferroelectrics. According to the first principles simulations[39,40], ferroelectric phase can be stabilized in PZO to removes the energetically costly interactions between head-to-tail dipoles while the long-range antiferroelectric order was broken. Thus, in this work, the PNRs may be formed to avoid the costly interactions in PZO as the long-range order was broken by implanted ions. The increase of polarization value may result from the coupling between ferroelectricity and lattice distortion[41] where the He implantation induced tetragonal phase has elongated c axis. Besides, the ion implantation gives rise to the increase of defect concentration, leading to high breakdown field strength in dielectric materials. These intrinsic point defects and eliminate shallow-hole trap states (such as isolated $Pb^{3+}$ and $V_{Pb}''$ defects with small activation energies) and form deep-level trap states[42-45], enabling the reduction of the leakage current (Fig. S6). What's more, the He bubbles observed in implanted PZO (Fig. S4) may induce local lattice distortion and trap carriers due to the high binding energy with some vacancy defects, which is beneficial for energy storage performance.

## III. CONCLUSIONS

In summary, we demonstrate He ion implantation as a powerful pathway to design PNRs, leading to the great enhancement of energy storage density in $PbZrO_3$ thin films. With the increase of implantation dose, the as-grown orthorhombic phase $PbZrO_3$ gradually transforms to tetragonal phase and even a partial of amorphous phase, resulting in the evolution from double hysteresis loops to single hysteresis loop which indicates the order (antiferroelectric)-disorder (relaxor-like ferroelectric) phase transition. This method enables the simultaneous enhancement of high-field polarizability and breakdown strength, tripling the energy storage density from 20.5 $J/cm^3$ to 62.3 $J/cm^3$. Our work opens up new possibilities for the enhancement of energy storage performance in dielectric capacitors.

## IV. EXPERIMENTAL METHODS

The LSMO layers were firstly grown on STO at 700 °C before the growth of PbZrO$_3$ layers at 560 °C using PLD. All the films were grown in an oxygen pressure of 15 Pa with a laser repetition rate of 8 Hz. After the deposition, all films were cooled down to room temperature at 10 °C/min in the oxygen atmosphere of 1000 Pa. Following the film growth, He ion implantation was carried out at 8 keV with a fluence from $2.5\times10^{14}$ to $5\times10^{15}$ ions/cm$^2$. A combination of X-ray diffraction (XRD), reciprocal space mappings (RSMs) and scanning transmission electron microscopy (STEM) were performed for structural characterizations. Morphology and ferroelectric domain structures were captured using atomic force microscopy (AFM) and piezoelectric force microscopy (PFM), respectively. To probe the electrical properties, Pt top electrodes with a diameter of 20 μm were deposited by magnetron sputtering. The P-E loops of Pt/PZO/LSMO devices were measured using a ferroelectric workstation (Precision Multiferroic, Radiant).

## SUPPLEMENTARY MATERIAL

**See the supplementary material for details on structure and property of** the as-grown and implanted PZO films.

## ACKNOWLEDGEMENTS

This work was supported by the National Natural Science Foundation of China (Grant Nos. 91963102 and U1832104), the Research Grants Council of Hong Kong (Project No. 15301421) and the Funding by Science and Technology Projects in Guangzhou (202201000008). Authors also acknowledge the financial support of Guangdong Science and Technology Project (Grant No. 2019A050510036), the Natural Science Foundation of Guangdong Province (Grant No. 2020A1515010736) and Guangdong Provincial Key Laboratory of Optical Information Materials and Technology (No. 2017B030301007). Y.G. thanks the funding from the State Key

Laboratory of Nuclear Physics and Technology, Peking University (No. NPT2019ZZ01). D.C. thanks the financial support from Department of Education of Guangdong Province (No. 2019KTSCX032), the Hong Kong Scholars Program (Grant No. XJ2019006) and the support of 2022 International (Regional) Cooperation and Exchange Programs of SCNU. S. Z. thanks the financial support by the German Research Foundation (Grant No. ZH 225/10-1).

## AUTHOR DECLARATIONS

**Conflict of Interest**

The authors have no conflicts to disclose.

**Author Contributions**

Y.L. and C.W. contributed equally to this work.

## DATA AVAILABILITY

The data that support the findings of this study are available from the corresponding author upon reasonable request.

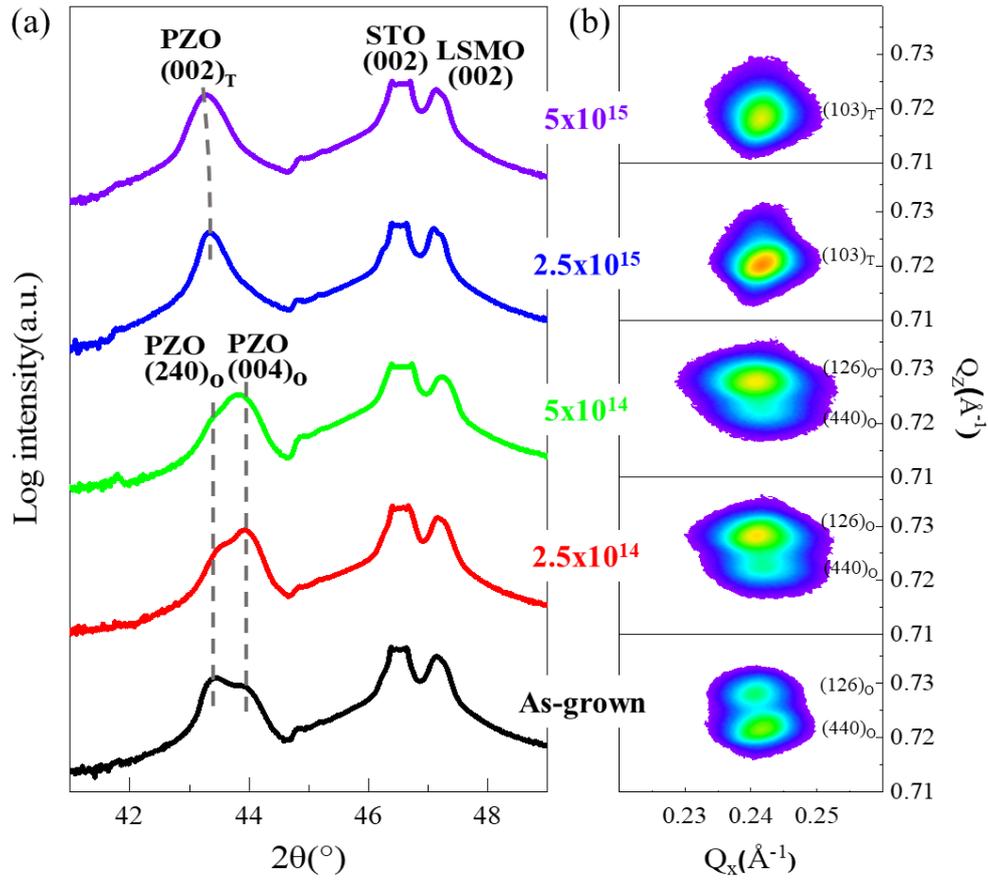

FIG. 1. Structural evolution of PbZrO$_3$ thin films with the increasing of He ion implantation doses. (a) X-ray $\theta$-$2\theta$ line scans and (b) Reciprocal space mapping studies around the (440)$_O$ diffraction peaks reveal the He ion implantation driven orthorhombic-tetragonal (O-T) phase transitions in PbZrO$_3$ thin films.

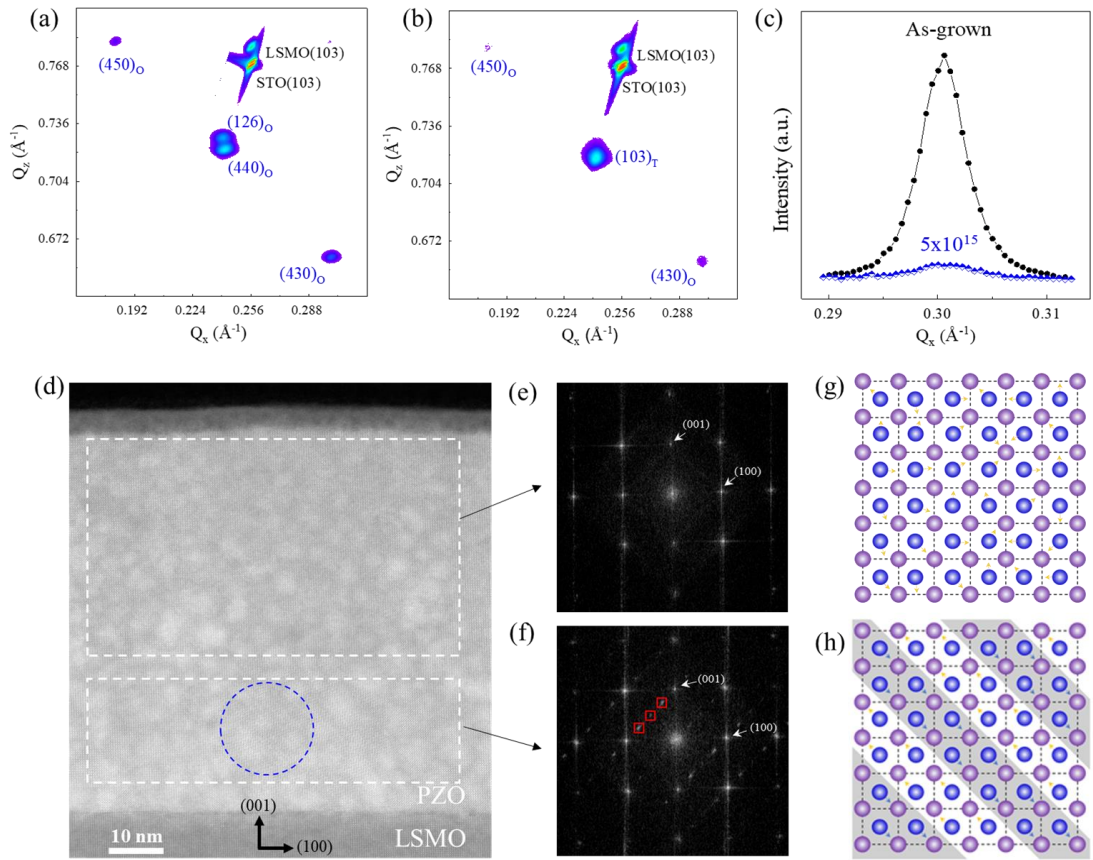

FIG. 2. Phase transition driven by ion implantation. Reciprocal space mappings study around the (103) reflection of (a) as-grown PZO and (b) implanted PZO with the dose of $5\times10^{15}$ ions/cm$^2$. (c) The (430)$_O$ diffraction peak intensity along the L-direction, which reveals strong Bragg diffraction intensity for the as-grown PZO but two orders of magnitude lower in the implanted PZO film. (d) The cross-sectional STEM image of implanted PZO sample with the dose of $2.5\times10^{15}$ ion/cm$^2$. (e) Fast Fourier transform (FFT) of the top dashed rectangle region of implanted PZO and (f) FFT of the region close to PZO/LSMO interface where the superstructure spots in red squares arise from the blue dash circled orthorhombic phase in Fig. 2(c). (g) Schematic of the implanted PZO unit cell with tetragonal phase and (h) schematic of PZO unit cell with the antiferroelectric structure.

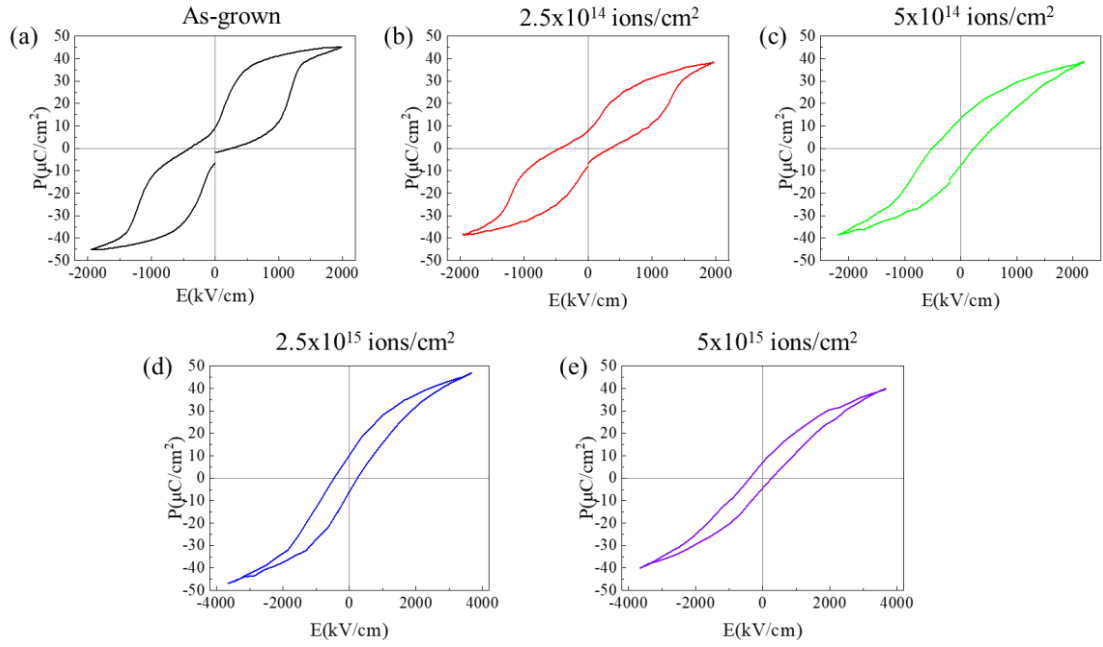

FIG. 3. Polarization - electric field hysteresis loops of the as-grown and implanted PZO films. (a) the as-grown sample and the implanted samples with the dose of (b) 2.5×10$^{14}$ ion/cm$^2$, (c) 5×10$^{14}$ ion/cm$^2$, (d) 2.5×10$^{15}$ ion/cm$^2$ and (e) 5×10$^{14}$ ion/cm$^2$.

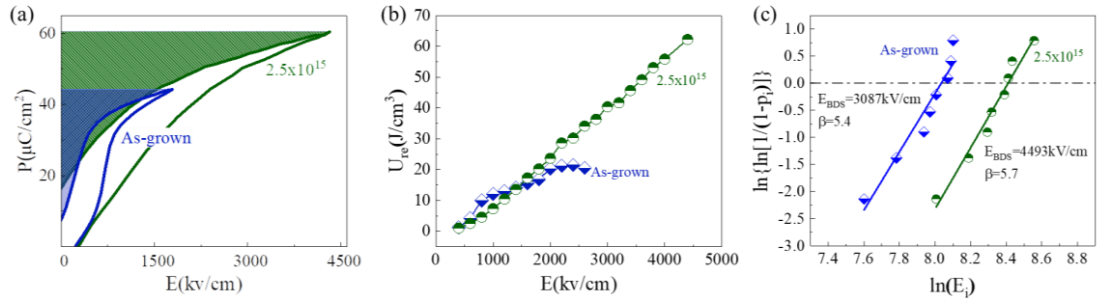

FIG. 4. Energy storage performance. (a) Unipolar hysteresis loops at maximum electric field measured at 10 kHz. (b) Energy density calculated from unipolar hysteresis loops. (c) Two-parameter Weibull distribution analysis of breakdown strengths.

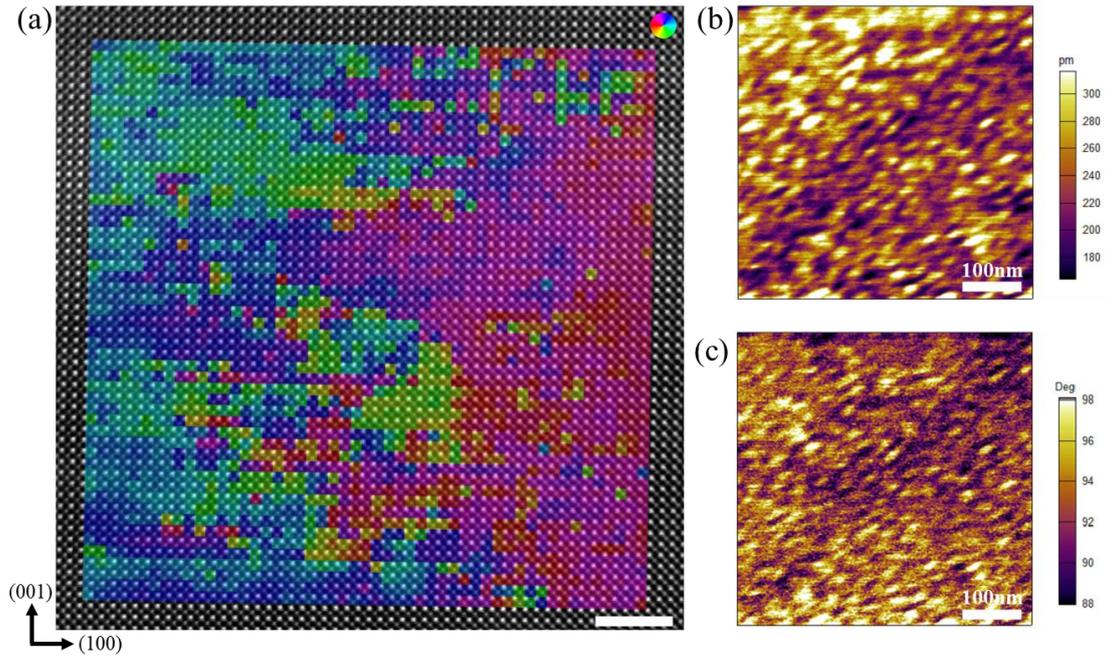

FIG. 5. Observation of polar nanoregions (PNRs) in the implanted PZO. (a) polarization vector mapping based on the STEM image of the implanted PZO with the dose of $2.5\times10^{15}$ ion/cm$^2$. The corresponding out-of-plane (b) amplitude and (c) phase PFM images of the implanted PZO film.